\documentclass [11pt,twoside]{article}

\usepackage{shrthnds}
\usepackage{cite}
\usepackage{graphicx,amsmath}
\usepackage{color}
\usepackage{subfigure}
\usepackage{setspace}

\usepackage{multirow}

\usepackage[hang,small]{caption}

\usepackage{geometry}
\usepackage{fancyhdr}
\usepackage{titlesec,titletoc}
  \titleformat{\section}{\Large\sf\bfseries}{\thesection}{1em}{}
  \titleformat{\subsection}{\large\sf\bfseries}{\thesubsection}{1em}{}

\title{\sf\bfseries 
Non-relativistic matter and Dark energy in a quantum conformal model
}

\author{\normalsize  
Pankaj Jain\footnote{email: pkjain@iitk.ac.in}~,
 Gopal Kashyap\footnote{email: gopal@iitk.ac.in}
and Subhadip Mitra\footnote{email: subhadipmitra@gmail.com}
}
\date{}

\begin{document}
\maketitle
\vspace{-0.6cm}
\bc
{\small Department of Physics, IIT Kanpur, Kanpur 208 016, India}
\ec

\centerline{\small\date{\today}}
\vspace{0.5cm}

\bc
\begin{minipage}{0.9\textwidth}\begin{spacing}{1}{\small {\bf Abstract:}
We consider a generalization of the standard model which respects 
quantum conformal invariance. This model leads to identically zero vacuum
energy. We show how non-relativistic matter and dark energy arises in
this model. Hence the model is shown to be consistent with observations.

}\end{spacing}\end{minipage}\ec

\vspace{0.5cm}\begin{spacing}{1.1}

\section{Introduction}
In a recent paper we considered a generalization of the standard model which
displays conformal invariance within the full quantum theory \cite{Jain:2015}. 
It has been shown earlier 
\cite{Englert:1976,JM,JMS,Shaposhnikov:2008a,Shaposhnikov:2008b,Singh:2011}
that it is possible to evade conformal anomaly \cite{Coleman:1971,Delbourgo,Capper1974,Duff1976,Dowker1976,Brown1977,Tsao1977,Fujikawa1980} if conformal
invariance is broken by a soft mechanism. 
In this case it is possible to use a dynamical scale  
for regularization which preserves conformal invariance.
For example, within the framework of dimensional regularization, 
the fixed mass scale $\mu$ is replaced by a real scalar field which we 
denote by the symbol $\chi$. 
This is called the GR-SI prescription \cite{Shaposhnikov:2008a}. 
The conformal symmetry 
may be broken spontaneously \cite{Englert:1976,Shaposhnikov:2008a,Shaposhnikov:2008b,Jain14} or by the background cosmic evolution \cite{JM,JMS}. 
The resulting non-zero classical value of $\chi$ 
provides the regularization mass scale.
Due to quantum conformal invariance, 
the trace of the energy momentum tensor $T_{\mu\nu}$ 
in this theory is found to be equal
to a total divergence, that is, 
\begin{equation}
T^\mu_\mu \sim \partial_\mu(\chi\partial^\mu\chi)
\label{eq:Tmumu}
\end{equation}
Hence its vacuum expectation value (VEV) is equal to zero. 
However its expectation value in other states need not be zero. 

Despite the fact that the theory respects quantum conformal invariance,
the problem related to the fine tuning of the cosmological constant
\cite{Weinberg,Padmanabhan}
remains in its simplest formulation. This has been discussed in 
detail in \cite{Jain:2015}. For example, the spontaneous breaking 
\cite{Wetterich} of
conformal invariance requires that we set one of the parameters in the
scalar potential to zero \cite{Shaposhnikov:2008a}. This parameter
is not protected by any symmetry and has to be set to zero order by
order in perturbation theory. Unless this parameter,
which we denote by the symbol $\lambda$, takes a very small value, 
the scalar field $\chi$ quickly decays to zero and the GR-SI prescription 
cannot be implemented \cite{Jain:2015}.  

In \cite{Jain:2015} we argued that this fine tuning problem can be evaded
if there exists another strongly coupled sector in the theory besides QCD. 
We considered a 
model in which the strongly coupled sector couples very weakly to
the standard model particles and hence provides candidates for dark matter
\cite{Meissner07,Meissner08,Hur11}. 
In this case the parameter $\lambda$ need not take a 
very small value and hence does not require fine tuning. We review 
this model later in this section. The main purpose of the present
paper is to show how non-relativistic matter and dark energy
can arise in this model.  
The presence of non-relativistic matter is not immediately 
obvious due to the condition displayed in Eq. \ref{eq:Tmumu}. The problem
is that on the right hand side we only obtain contributions from 
the scalar fields in the theory. In Eq. \ref{eq:Tmumu} we have displayed
only one such term. Analogous terms are present for other scalar fields. 
However we also require contributions from fermionic fields, such as, 
protons and neutrons. These do not appear explicitly in $T^\mu_\mu$,
while we expect them to contribute. Hence it is not clear 
whether the implications of quantum conformal invariance, i.e. Eq. 
\ref{eq:Tmumu}, are consistent with observations, in particular,
solar system physics and cosmology. As we shall show in this paper,
the conformal theory necessarily leads to additional contributions
besides non-relativistic matter. This means that it is not possible to 
only have non-relativistic matter which is permissible within the
standard framework. In the present paper we examine these additional 
contributions. These might lead to interesting signals in astrophysics
and cosmology. However in the present paper we shall primarily 
be interested in demonstrating that 
it is possible to suppress these 
contributions by a suitable choice of parameters. Hence we argue that the
model provides a consistent framework for cosmology, free from 
the fine tuning problem of the cosmological constant. The model also
does not suffer from the problem of fine tuning of the Higgs mass
due to conformal invariance \cite{Shaposhnikov:2008a}.   

\subsection{Review of the conformal model}
In this subsection 
we briefly review the conformal model described in \cite{Jain:2015}. 
Here we shall directly discuss the generalization of the standard model
rather than the toy model considered in \cite{Jain:2015}. We display the 
action directly in $d$-dimensions. The action can be written as
\begin{equation}
{\cal S} = {\mc S}_G  + {\mc S}_{SM} + {\mc S}_{D} 
\label{eq:LagrangianSM}
\end{equation}
where ${\mc S}_G$ is the gravitational action, ${\mc S}_{SM}$ 
represents the conformal extension of the standard model and
${\mc S}_{D}$ represents the strongly coupled dark sector. 
The gravitational action can be expressed as,
\begin{equation}
{\mc S}_G = \int d^dx\sqrt{-g} \left( {M_{PL}^2\over 16\pi} R -{\xi\over 2}\chi^2 R  \right)
\label{eq:SG2}
\end{equation}
where the first term is the standard Einstein action and second term
represents a non-minimal coupling of the real scalar field
 $\chi$ with gravity \cite{CCJ,Deser}. 
Here $M_{PL}$ denotes the Planck mass and $\xi$ is an additional parameter.
We may add 
similar terms for other scalar fields. In Ref. \cite{CCJ} it has been suggested
that we should set $\xi=1/6$ since it leads to improved energy momentum tensor. Furthermore
in a conformal model it leads to $R=T^\mu_\mu=0$, instead of Eq. 
\ref{eq:Tmumu}. This also holds in $d$ dimensions if we choose,  
\begin{equation}
\xi= {(d-2)\over 4(d-1)}\, .
\label{eq:confcoup1}
\end{equation}
In our analysis we shall simply set this parameter to zero. A non-zero 
value of this parameter would be useful for a detailed fit to the cosmological 
observations. However it is not essential for our analysis and does not
affect our conclusions as long as it is different from the value given
in Eq. \ref{eq:confcoup1}.

The action for the
 conformal generalization of the standard model can be expressed as
\ba
\mathcal{S}_{SM} &=& \int d^dx \sqrt{-g}\Bigg[{1\over 2}g^{\mu\nu}\partial_\mu\chi
\partial_\nu\chi+ 
 g^{\mu\nu} (D_\mu \mc H)^\dag(D_\nu \mc H) -\frac14
g^{\mu\nu} g^{\alpha\beta}(\mathcal{A}^i_{\mu\alpha} \mathcal{A}^i_{\nu\beta}
\nn\\
&+& \mathcal{B}_{\mu\alpha} \mathcal{B}_{\nu\beta} + \mathcal{G}^a_{\mu\alpha}
 \mathcal{G}^a_{\nu\beta}) (\zeta^2)^{\delta}
 - {\lambda_1\over 4}   (2\mc H^\dag \mc H-\lambda_2\chi^2)^2 
(\zeta^2)^{-\delta}-{\lambda\over 4}\chi^4(\zeta^2)^{-\delta}  \Bigg]\nonumber\\
 &+& {\cal S}_{\rm fermions},
\label{eq:S_EW_d}
\ea
where $\delta = (d-4)/(d-2)$,
$\mc H$ is the Higgs multiplet and $\mc{G}^a_{\mu\nu}$, $\mc{A}^i_{\mu\nu}$ and
$\mc{B}_{\mu\nu}$ denote the SU(3), SU(2) and U(1) field strength tensors. 
Here 
\begin{equation}
\zeta^2 =\beta\chi^2+2\beta_1 \mc H^\dag \mc H
\label{eq:zeta}
\end{equation}
and $\beta$, $\beta_1$ are parameters. 
For simplicity we may set $\beta=1$ and $\beta_1=0$ which leads to $\zeta^2=\chi^2$. 
Besides the standard model fields, 
the only additional field in this 
action is the real scalar field $\chi$.  
The action, ${\cal S}_{\rm fermions}$, is given by
\ba
{\mc S}_{\rm fermions} &=&\int d^d x\, e\, \left({\overline\Psi}_{\rm L}i
\gamma^\mu  {\cal D_\mu} \Psi_{\rm L} +
{\overline\Psi}_{\rm R}i \gamma^\mu  {\cal D_\mu} \Psi_{\rm R}  \right)\nn\\
&&- \int d^dx\, e\, (g_Y\overline{\Psi}_{\rm L} {\mc H}\Psi_{\rm R} 
(\zeta^2)^{- \delta/2} + h.c.),
\label{eq:Sfermions}
\ea
where, as usual, 
$\Psi_L$ and $\Psi_R$ are the left and right handed projections
of a fermion field and the Yukawa coupling is denoted by
$g_Y$. Here we have displayed
the action for only one left handed SU(2) doublet and a right handed SU(2) 
singlet. Similar terms can be added for all the fermions. 
Furthermore,
$e={\rm det}(e_\mu^{~a})$, 
 and  $e_\mu^{~a}$ is the
vielbein. 
The Higgs multiplet can be decomposed as
\begin{equation}
\mc H = {1\over \sqrt{2}} \left(\begin{array}{c}\phi_1+i\phi_2\\
 \phi_3+i\phi_4 \end{array}\right) 
\label{eq:Higgsmultiplet}
\end{equation}

The strongly coupled dark matter action can be expressed as 
\cite{Meissner07,Meissner08,Hur11}, 
\begin{equation}
{\mc S}_D = \int d^dx\sqrt{-g} \left[ -{1\over 4} G_{\mu\nu}^a G^{a\mu\nu }
\left(\zeta^2\right)^\delta 
+i\bar\xi^i \gamma^\mu D_\mu\xi^i - g_Y\bar\xi^i\chi\xi^i\left(\zeta^2\right)^{(-\delta/2)}  \right]
\label{eq:stronglycoupled}
\end{equation}
where $G_{\mu\nu}^a$ is the field strength tensor and $\xi^i$ represent
fermion fields. We refer to this strongly coupled sector as hypercolor.
Here we have included only one multiplet of hypercolor fermions. In general
several multiplets might exist. This strong sector couples to the electroweak
sector only through its coupling to the scalar field $\chi$ and terms such
as $(\zeta^2)^\delta$ . 
The field $\chi$ couples to the electroweak sector by its interaction with
the Higgs field. This interaction is proportional to $\lambda_2$. As
discussed in \cite{Jain:2015} we expect this coupling to be very small.
The terms such as $(\zeta^2)^\delta$ contribute only at loop orders. 
As discussed in \cite{Jain:2015},
these loop contributions are suppressed by Planck mass and hence are very
small. 
The matter action, i.e. $\mc S_{SM}+\mc S_D$, 
in $d$ dimensions is symmetric under the conformal transformation, 
\begin{equation}
\Phi\rightarrow {\Phi\over \Omega}\, ,g_{\mu\nu}\rightarrow
\Omega^{\,b} g_{\mu\nu}\,, 
 A_\mu\rightarrow A_\mu\,, \Psi\rightarrow \Psi/\Omega^{\,c}
\label{eq:globalconftrans}
\end{equation} 
where $b=4/(d-2)$,  $c=(d-1)/(d-2)$, $\Phi$ is a scalar field, 
$\Psi$ a fermion field and $A_\mu$
a vector field.

We need to break conformal symmetry by a soft mechanism. The relevant equations are
the classical equations of motion for the Higgs field and $\chi$ and
the dynamical equations for the strongly coupled sector. These equations
for the strongly coupled sector are not well understood. However we
know that these generate the condensates for the gauge fields and
fermions. In making quantum computations, we need to regulate the
action. In our case this is accomplished by introducing the dynamical
field $\zeta$. The classical value of this field is determined self-consistently
by solving the classical equations of motion for $\chi$ and $\mc H$. 
In the Higgs multiplet we set $\phi_1=\phi_2=\phi_4=0$ and focus
on the physical Higgs field, $\phi_3$, which is expected to have a
non-zero VEV. The classical equations of motion of $\phi_3$ and $\chi$ 
can be written as,  
\begin{eqnarray}
g^{\mu\nu}\phi_{3;\mu;\nu} + \lambda_1\phi_3
(\phi_3^2-\lambda_2\chi^2)  = 0 \nonumber\\
g^{\mu\nu}\chi_{;\mu;\nu} +\lambda\chi^3 - \lambda_1\lambda_2 \chi
(\phi_3^2-\lambda_2\chi^2) + g_1\langle\bar\xi^i\xi^i\rangle = 0
\end{eqnarray}
where we have replaced $\bar\xi^i\xi^i$ by its vacuum expectation value
$\langle\bar\xi^i\xi^i\rangle$. The strong interaction dynamics leads to a
non-zero value of this condensate. We express this as,
\begin{equation}
\langle \bar\xi^i\xi^i\rangle =\Lambda_S^3\,. 
\end{equation}
Once this condensate is generated we determine space-time independent
solution to the equations of motion for 
$\chi$ and $\phi_3$.
 The resulting solution can be expressed as,
\begin{eqnarray}
v = \sqrt{\lambda_2}\, \eta\nonumber\\ 
\lambda\eta^3 = - g_1\langle\bar\xi^i\xi^i\rangle 
\label{eq:solutionveta}
\end{eqnarray}
where $v$ and $\eta$ are the vacuum expectation values of the fields $\phi_3$
and $\chi$ respectively. 
Hence these leads to the electroweak scale with a suitable choice of the 
parameters $\lambda_2$ and $\eta$. 
The parameters $\lambda_2$ and $\lambda$ are expected to take very small 
values in this model. However it has been shown in \cite{Jain:2015}
that no fine tuning of these parameters is required at loop orders. 

An important point about this model is that the curvature scalar $R$ is 
 proportional to a total derivative. The precise value depends on the
parameter $\xi$. For simplicity, here we set $\xi=0$ although we do not
need to make this choice. In general, the parameter $\xi$ may be useful
for a detailed cosmological fit to data.
For $\xi=0$, we obtain
\begin{equation}
{R\over 8\pi G} = T^\mu_\mu = -(\chi\partial^\mu \chi)_{;\mu} + ...
\label{eq:RTmumu}
\end{equation}
where the dots indicate that similar contributions are obtained
from all scalar fields in the theory. 
We point out that for $\xi\ne 0$, this equation in $d$ dimensions
becomes
\begin{equation}
{R\over 8\pi G} = T^\mu_\mu = -\left[1-4\xi\left({d-1\over d-2}\right) \right]
(\chi\partial^\mu \chi)_{;\mu} + ...
\label{eq:RTmumu1}
\end{equation}
The VEV of the terms on the
right hand side of this equation is zero since these are total derivatives.
Hence these terms do not contribute to vacuum energy.  
We expect that $R$ should get a contribution from non-relativistic 
matter. This is not obvious from Eq. \ref{eq:RTmumu} in which only the
scalar fields contribute and there are no contribution from fermions, i.e.
 physical
fields such as protons, neutrons and electrons. 
At current time,
we expect that the contribution to this equation from 
massive scalar fields, such as the Higgs field, would be negligible. 
Only the fields which have very low mass may contribute. In next 
section we investigate whether such contributions can effectively lead
to non-relativistic matter.

\section{Non-relativistic Matter}
\label{sec:NRmatter}
In the previous section we have outlined the main problem with the
model. It is not clear how non-relativistic matter arises in this case. 
The basic problem is illustrated by Eq. \ref{eq:RTmumu} where only
scalar fields contribute to the trace of the energy momentum tensor, whereas
we also expect contributions from fermion fields such as, protons, neutrons
and electrons. In this section we study such contributions in more detail.
The dominant contribution to the mass of visible matter is given by
protons and neutrons. Within our framework, we face a problem in handling
these particles due to their composite nature. We handle them by 
introducing an effective interaction term of protons and neutrons 
with the Higgs field. These terms are the same as those for up and down
quarks with an effective interaction which can be expressed in terms of
a form factor. At zero momentum transfer we expect this form factor to 
be proportional to the mass of these particles. Hence the effective couplings
for proton and neutron are $g_p=m_p/v$ and $g_n=m_n/v$ respectively
where $m_p$ and $m_n$ are the corresponding masses. We denote these
fields by the symbols $\psi_p$ and $\psi_n$ respectively. Their 
effective Yukawa interaction terms can be expressed as 
\begin{equation}
\mc L_Y = -g_p\bar\psi_p\psi_p \phi_3 - g_n\bar\psi_n\psi_n \phi_3 
\end{equation}

The equations of motion for the scalar fields, including contributions from 
the Yukawa terms, can be written as, 
\begin{eqnarray}
g^{\mu\nu}\phi_{3;\mu;\nu} + \lambda_1\phi_3
(\phi_3^2-\lambda_2\chi^2)  +g_p\bar\psi_p\psi_p
+ g_n\bar\psi_n\psi_n
 = 0 \nonumber\\
g^{\mu\nu}\chi_{;\mu;\nu} +\lambda\chi^3 - \lambda_1\lambda_2 \chi
(\phi_3^2-\lambda_2\chi^2) + g_1\langle\bar\xi^i\xi^i\rangle = 0
\label{eq:eomrho}
\end{eqnarray}
We are interested in the contributions of protons and neutrons which
act as non-relativistic matter. We can replace the fermion bilinears
in terms of their energy density, i.e.,
\begin{equation}
m_p \bar\psi_p\psi_p + m_n \bar\psi_n\psi_n = \rho
\end{equation}
where $\rho$ is the non-relativistic energy density. 
If we set $\rho=0$ the solution to Eq. \ref{eq:eomrho} is given by
Eq. \ref{eq:solutionveta}. However in the presence of $\rho$, 
which is treated as a small perturbation, we
expect that the solution can be expressed as,
\begin{eqnarray}
\phi_3 &=& v + \delta\phi_3\nonumber\\
\chi &=& \eta + \delta\chi
\end{eqnarray}
where $\delta\phi_3 $ and $\delta\chi$ are small and determined by $\rho$. 

Let us now set the
space and time derivative of the Higgs field to zero. We expect this 
to be a good approximation since this field is relatively heavy and
the potential terms dominate in the equation of motion. The derivatives
are related to the derivatives of $\rho$ and hence expected to be 
negligible. We shall explicitly show
later that this is a good approximation. The $\phi_3$
equation of motion then yields
\begin{equation}
\lambda_1(\phi_3^2-\lambda_2\chi^2) = -{1\over \phi_3}{\rho\over v}
\approx -{\rho\over v^2}
\label{eq:pot1}
\end{equation}
Substituting this into the equation of motion for $\chi$ we obtain
\begin{equation}
g^{\mu\nu}\chi_{;\mu;\nu} +\lambda\chi^3 
+ g_1\langle\bar\xi^i\xi^i\rangle + {\lambda_2\chi\over v^2}\rho = 0
\label{eq:eomchi11}
\end{equation}
The term proportional to $\rho$ acts as a small perturbation in this
equation which determines the deviation of $\chi$ from its VEV $\eta$
which is a constant. 
Since $\rho$ is small
we can replace $\chi$ in the last term in Eq. \ref{eq:eomchi11} by $\eta$.
We next need to determine the expectation value $\langle\bar\xi^i\xi^i\rangle$
when $\rho\ne 0$. Recall that if $\rho=0$ then 
$\langle\bar\xi^i\xi^i\rangle$ is given by Eq. \ref{eq:solutionveta}.
However in the present case the background value of $\chi$ will
be slightly different from $\eta$ which will also lead to a shift
in $\langle\bar\xi^i\xi^i\rangle$. 
Computing this shift, however, is difficult since this requires us to
perturb the equations corresponding to the strong interaction dynamics. 
Here
we make a simple approximation, 
\begin{equation}
\lambda\chi^3 + g_1 \langle\bar\xi^i\xi^i\rangle = 0
\label{eq:chiapprox}
\end{equation} 
i.e., we simply replace $\eta$ in Eq. \ref{eq:eomchi11} by the modified
background value of $\chi$. 
Below we give some justification for this approximation.
Under this approximation the equation of motion of $\chi$ yields
\begin{equation}
g^{\mu\nu}\delta\chi_{;\mu;\nu} + {\rho\over \eta} = 0
\label{eq:eomdeltachi}
\end{equation} 
i.e. $\chi$ (or $\delta\chi$) acts as a massless field whose evolution is controlled by
the non-relativistic energy density. 
Our approximation, Eq. \ref{eq:chiapprox}, is partially justified 
by the fact that we expect a massless scalar field in our theory due to
soft breaking of conformal symmetry. If we use Eq. \ref{eq:chiapprox}, we
find that this field is $\chi$. If Eq. \ref{eq:chiapprox} is not a 
good approximation, then this field would be a linear combination of $\chi$,
$\phi_3$ and a scalar bound state of the dark fermions. We again
expect an equation analogous to Eq. \ref{eq:eomdeltachi} for the resulting
massless field. Hence we do not expect the physical consequences to
be modified significantly even if Eq. \ref{eq:chiapprox} is 
not reliable.

We may also consider the case in which the strongly interacting dark sector
is absent. In this case we need to arbitrarily set $\lambda=0$ or 
extremely small. Hence this model requires fine tuning. However the model
still satisfies conformal invariance and obeys Eq. \ref{eq:RTmumu}. Hence
the problem we are trying to address is also present. Due to the absence
of dark strong sector this model is easier to handle. In this case we find that
Eq. \ref{eq:chiapprox} is trivially satisfied since $\lambda=0$. Hence
we again obtain Eq. \ref{eq:eomdeltachi}. The remaining analysis in this
model is same as presented in the remainder of this paper for the model 
with $\lambda\ne 0$.  

Yet another possibility is set $\lambda$ to be extremely small but not
zero. In this case the scalar field evolves with time. As discussed in
\cite{Jain:2015}, we can choose $\eta$, the classical value of $\chi$,
to be sufficiently large so that it does not decay to zero over
the lifetime of the Universe. It turns out that this is possible only
if $\eta>>M_{PL}$. An explicit calculation shows that this model 
with $\xi=0$ produces a very large energy density which is cosmologically
unacceptable. If we instead 
choose the value given in Eq. \ref{eq:confcoup1} 
then the energy density is zero. Hence it is clear that we can produce
cosmologically acceptable energy density if we set $\xi$ very close
to the value in Eq. \ref{eq:confcoup1}. However this will require fine
tuning of this coupling. It is not clear if this fine tuning persists 
at loop orders since the loop corrections are suppressed by Planck mass
and expected to be very small. In any case we do not pursue this 
possibility in the present paper.

We next determine the derivatives of $\phi_3$ in order to check if we
were justified in ignoring those. We have
\begin{equation}
\phi_3^2 = \lambda_2\chi^2 - {\rho\over v^2\lambda_1}
\end{equation} 
which leads to
\begin{equation}
\delta\phi_3 = {\lambda_2\eta\over v}\delta\chi - {\rho\over 2v^3\lambda_1}
\end{equation} 
We next compute $g^{\mu\nu}\delta\phi_{3;\mu;\nu}$. By using Eq. 
\ref{eq:eomdeltachi}, the second derivatives of $\phi_3$ can expressed in
terms of the second derivatives of $\rho$. For a slowly varying 
$\rho$ it is easy to check that $g^{\mu\nu}\delta\phi_{3;\mu;\nu}$ is 
negligible compared to the terms we kept in the equation of 
motion of $\phi_3$. 

\subsection{Time Independent Energy Density}
Let us next consider the case of a localized time independent energy 
density. The equation of motion of $\chi$ in this case can be expressed
as 
\begin{equation}
-\nabla^2\delta\chi+ {\rho\over \eta} = 0
\label{eq:deltachirho}
\end{equation} 
 Hence we see that the scalar field 
$\chi$ provides a new force which couples to mass with an effective
gravitational constant equal to $1/\eta^2$. 

We next determine the contribution to the Einstein's equations.   
The energy-momentum tensor can be written as
\begin{equation}
T_{\mu\nu} = \partial_\mu\phi_3\partial_\nu\phi_3 + \partial_\mu\chi
\partial_\nu\chi +
\bar \psi_p i\gamma_\mu\partial_\nu\psi_p
+\bar \psi_n i\gamma_\mu\partial_\nu\psi_n
+ \bar \xi^i i\gamma_\mu D_\nu\xi^i
- g_{\mu\nu} \mc L 
\end{equation}
We shall drop the derivatives of the Higgs field.  
In the present case the time derivative of $\chi$ is also zero.
The resulting $0-0$ component of the energy-momentum tensor can be expressed as
\begin{equation}
T_{00} =  
\bar \psi_p i\gamma_0\partial_0\psi_p + 
\bar \psi_n i\gamma_0\partial_0\psi_n + 
\langle\bar\xi^i i\gamma_0 D_0\xi^i\rangle -g_{00} \left[-{1\over 2}(\nabla\chi)^2 - {\lambda\over 4}\chi^4 - {\lambda_1\over 4}(\phi_3^2- \lambda_2\chi^2)^2\right]
\end{equation}
Here we use the FRW metric and $g_{00}=1$. 
We have also used the fermion equations of motion in order to simplify this 
expression.
Using Eq. \ref{eq:pot1} and the fact that $\lambda_1\sim 1$ we find that 
$(\phi_3^2- \lambda_2\chi^2)^2\sim \rho (\rho/v^4)$. Hence in most 
cases of observable energy density, this term is negligible compared to $\rho$
and we shall drop it. 
The equation of motion for the fermion field $\xi$ implies
\begin{equation}
g^{\mu\nu}\bar\xi^i i\gamma_\mu D_\nu\xi^i - g_1\chi\bar\xi^i\xi^i = 0
\label{eq:eomxi}
\end{equation}
We expect that the vacuum expectation values satisfy
\begin{equation}
\langle\bar\xi^i i\gamma^0 D_0\xi^i\rangle 
= \langle\bar\xi^i i\gamma^1 D_1\xi^i\rangle  
= \langle\bar\xi^i i\gamma^2 D_2\xi^i\rangle  
= \langle\bar\xi^i i\gamma^3 D_3\xi^i\rangle  
\end{equation}
This is because all components should be equal in vacuum. 
Hence by using Eq. \ref{eq:eomxi} we obtain
\begin{equation}
\langle\bar\xi^i i\gamma^0 D_0\xi^i\rangle  = {1\over 4} g_1\chi
\langle\bar\xi^i\xi^i\rangle
\label{eq:eomxi1}
\end{equation}
We point out that, for simplicity, here we perform the analysis in 
4 dimensions but the entire calculation goes through also in $d$ dimensions. 
Substituting the above equation into $T_{00}$ we obtain
\begin{equation}
T_{00} = 
\rho + {1\over 2}(\nabla\chi)^2 
\end{equation}
where we have set
\begin{equation}
\bar \psi_p i\gamma_0\partial_0\psi_p + 
\bar \psi_n i\gamma_0\partial_0\psi_n 
= m_p\bar \psi_p \psi_p
+ m_n\bar \psi_n \psi_n
=\rho 
\label{eq:NRrho}
\end{equation}
and we can also replace $\chi$ by $\delta\chi$.

We next estimate the field $\delta\chi$ for a spherically symmetric  
distribution $\rho(r)$. Using Eq. \ref{eq:deltachirho} we obtain
\begin{equation}
\vec \nabla\chi = \vec\nabla\delta\chi = {M\over 4\pi\eta}{\hat r\over r^2}
\end{equation}
where $M$ is the total mass contained within radius $r$. Hence we find that
\begin{equation}
T_{00} = \rho + {M^2\over 2(4\pi)^2\eta^2r^4}
\end{equation}
The first term is the standard non-relativistic matter. The second
term is the extra term that comes along in our model. In order to 
estimate the relative importance of the two terms we consider the
gravitational potential of Sun at Earth. We integrate the two terms.
The integral over $\rho$ leads to the total mass $M$ of the Sun. The integral
over the second term is equal to 
$$M \left[{M\over 8\pi\eta^2 R}\right]\, .$$
Hence the relative importance of the second term is determined by the
expression inside the brackets. This is found to be equal to
\begin{equation}
{M\over 8\pi\eta^2 R} \sim 10^{-9} \left[{1 AU\over R}\right]\left[{M_{PL}\over
\eta}\right]^2
\end{equation} 
Hence the second term is small but not negligible if $\eta =M_{PL}$. 
However we can choose the parameter $\eta$ to be sufficiently large
so that this term does not lead to disagreement with experimental 
data on the solar system scale. 
We next estimate the space-space components of the energy-momentum tensor.
We obtain
\begin{equation}
T_{ij} = \partial_i\chi\partial_j\chi + {1\over 2}g_{ij}(\nabla\chi)^2
\end{equation}
It is clear that both terms are of the same order of magnitude as the
extra term in $T_{00}$. Hence by a suitable choice of $\eta$ these 
can be made sufficiently small. 
Similarly the additional force 
provided by the field $\chi$, as given in Eq. \ref{eq:deltachirho}, 
can also be suppressed to the required value by choosing a sufficiently
large $\eta$. Hence with a suitable choice of $\eta$ our 
model is consistent with physics on the solar system and smaller
distance scales.    

\subsection{Cosmic Evolution}
We next examine the contribution of non-relativistic matter to cosmic
evolution. As in the previous subsection, the non-relativistic matter
will be dominated by protons and neutrons. The main point is that 
in our formalism there are necessarily additional contributions to the
Einstein's equations besides the standard contribution due to 
 non-relativistic matter. Here we examine these contributions 
in order to determine if they are sufficiently small in some limit. 

In the present case, Eq. \ref{eq:eomdeltachi} leads to
\begin{equation}
{d^2\delta\chi\over dt^2} + 3H {d\delta\chi\over dt} = -{\rho\over \eta}
\label{eq:eomdeltachi1}
\end{equation} 
The components of the energy-momentum tensor are given by
\begin{equation}
T_{00} = \dot\chi^2 
+\bar \psi_p i\gamma_0\partial_0\psi_p 
+\bar \psi_n i\gamma_0\partial_0\psi_n 
+\langle \bar \xi^i i\gamma_0 D_0\xi^i \rangle
-g_{00}\left[{\dot\chi^2\over 2}-\lambda\chi^4\right] = {\dot\chi^2\over 2} + \rho
\end{equation}
where we have used Eqs. \ref{eq:chiapprox}, \ref{eq:eomxi} and \ref{eq:eomxi1}.
We have also used the equations of motion for
$\psi_p$ and $\psi_n$ and Eq. \ref{eq:NRrho}.  
We also obtain
\begin{equation}
T^i_j = \langle \bar \xi^i i\gamma^i D_j\xi^i \rangle
-\delta^i_j \left[{\dot\chi^2\over 2}-\lambda\chi^4\right] = -\delta^i_j\, {\dot\chi^2\over 2} 
\end{equation}
Hence we see that we get the standard contribution from non-relativistic 
matter along with an extra term proportional to $\dot\chi^2$ or
equivalently $(d\delta\chi/dt)^2$. We next
evaluate the contribution of this term to cosmic evolution by using
Eq. \ref{eq:eomdeltachi1}. We first ignore the second derivative of $\delta\chi$. This leads to 
\begin{equation}
{d\delta\chi\over dt} \sim {\rho\over \eta H}
\end{equation}
Here $\rho$ is the non-relativistic energy density. Let us consider the 
phase in which this dominates the cosmic energy density. Here $\rho$ 
represents the contribution only of the visible matter. We may assume 
that the dark matter also gives a similar contribution and add
its contribution to $\rho$.  However a detailed
evaluation requires an explicit model of dark matter in our framework. 
Here we shall not go into these details and focus only on visible matter. 
Assuming that the non-relativistic matter 
dominates, we obtain
\begin{equation}
\rho\sim M_{PL}^2 H^2
\end{equation}
This implies that
\begin{equation}
\left({d\delta\chi\over dt}\right)^2 \sim \rho\, \left({\rho\over \eta^2 H^2}
\right)\sim \rho \left({M_{PL}^2\over \eta^2}\right)
\label{eq:chidotoverrho}
\end{equation}
This shows that the additional contribution is suppressed by a factor 
$(M_{PL}/\eta)^2$ and is negligible in the limit $\eta>> M_{PL}$.

We next consider solution to Eq. \ref{eq:eomdeltachi1} without neglecting
any term. We first assume that non-relativistic matter dominates 
cosmic energy density. 
In this case the scale factor is given by, $a(t) \propto t^{2/3}$, $H=2/3t$ and
$\rho= \rho_0 t_0^2/t^2$. Here $\rho_0$ and $t_0$ represent the current
density and time respectively.  
We seek a solution which will go to zero as $\rho_0\rightarrow 0$. 
Such a solution can be expressed as, 
\begin{equation}
\delta\chi = \delta\chi_0\, \ln(t_0/t) 
\end{equation} 
Substituting this into
Eq. \ref{eq:eomdeltachi1}, we obtain, $\delta\chi_0=\rho_0 t_0^2/\eta$.
Hence we find
\begin{equation}
{d\delta\chi\over dt} = {\rho_0 t_0^2\over \eta t}
\end{equation} 
The extra term in the energy momentum tensor is proportional to the square of
this term. Comparing this with $\rho$ we obtain
\begin{equation}
{1\over \rho} \left({d\delta\chi\over dt}\right)^2 \sim 
{\rho_0 t_0^2\over \eta^2}\sim {H_0^2M_{PL}^2\over \eta^2}{1\over H^2_0}
\sim {M_{PL}^2\over \eta^2} 
\end{equation}
which is same as that obtained in Eq. \ref{eq:chidotoverrho}. Hence 
this leads to the same constraint as obtained earlier.  
The analysis for the case of radiation domination, such that, $a\propto t^{1/2}$, and $H=1/2t$ is similar and leads to the same conclusion. For the 
case of vacuum domination, $a(t)=a_0e^{H(t-t_0)}$ we seek an approximate
solution of the form
\begin{equation}
\delta \chi = A(t) e^{-3Ht}
\end{equation}
where $A(t)$ is assumed to be a slowly varying function of $t$ such that
we can neglect its second derivative. We find that the solution is such
that
\begin{equation}
{dA\over dt} = {\rho_0e^{3Ht_0}\over 3H\eta}
\end{equation}
which is a constant. 
This leads to 
\begin{equation}
{d\delta \chi\over dt} = (-3HA+\dot A)\left({a_0\over a}\right)^3 e^{-3Ht_0} 
= \left[-{t\over \eta} + {1\over 3H\eta}\right]\rho \sim {\rho\over H\eta}
\end{equation}
where we have set $t\sim 1/H$. This again leads to the same result as
given in Eq. \ref{eq:chidotoverrho}.

To summarize, we find that in all cases  
the additional contributions to the energy-momentum tensor are 
suppressed by the factor $(M_{PL}/\eta)^2$. Hence we see that by a 
suitable choice of parameters we obtain the standard cosmic evolution. 
It is of course of considerable interest to determine the change 
in cosmic evolution due to the additional contributions in order to 
test if there is any evidence for them in data. However we postpone this 
study to a future paper. 

\subsection{Dark Energy}
The introduction of dark energy in our framework is straightforward. We 
simply add a cosmological constant term. In our framework the vacuum energy is 
identically zero and hence such a term can be added 
without requiring any fine tuning.

\section{Conclusions}
We have analysed a model in which the matter sector displays quantum 
conformal invariance. The trace of the energy momentum tensor is
found to be equal to a total divergence. A useful feature of the model
is that it leads to zero vacuum energy density. 
Hence we can add a small cosmological constant term which does not require
any fine tuning due to quantum corrections from the matter sector.
The model contains a 
strongly interacting dark matter sector. The
conformal symmetry is broken by the strong interaction dynamics in this
sector which leads to formation of condensates. The model is free from
the fine tuning problem of the cosmological constant. However 
the model does not admit non-relativistic matter in the standard manner. 
Such a contribution is necessarily associated with an additional 
contribution from the scalar fields in the model. In this paper
we have considered such additional contributions and have shown that
these are small in the limit when the classical value of the scalar
field $\chi$ is much larger than the Planck mass. Hence we argue that
the model is consistent with observations.   
A detailed fit of the model to cosmological data is postponed to future 
work.

\bigskip
\noindent
{\bf \large Acknowledgements:}  Gopal Kashyap thanks the Council
of Scientific and Industrial Research (CSIR), India for providing his Ph.D.
fellowship.

\end{spacing}
\begin{spacing}{1}
\begin{small}

\end{small}
\end{spacing}
\end{document}